\documentclass[sn-basic]{sn-jnl}

\usepackage{graphicx}%
\usepackage{multirow}%
\usepackage{amsmath,amssymb,amsfonts}%
\usepackage{amsthm}%
\usepackage{mathrsfs}%
\usepackage[title]{appendix}%
\usepackage{xcolor}%
\usepackage{textcomp}%
\usepackage{manyfoot}%
\usepackage{booktabs}%
\usepackage{algorithm}%
\usepackage{algorithmicx}%
\usepackage{algpseudocode}%
\usepackage{listings}%
\usepackage{hyperref}
\usepackage{ulem}
\hypersetup{
citecolor= blue, 
    colorlinks=true,
    linkcolor=blue, 
    filecolor=magenta,      
    urlcolor=magenta,
    pdftitle={Overleaf Example},
    pdfpagemode=FullScreen,
    }

\theoremstyle{thmstyleone}%
\theoremstyle{thmstyletwo}%

\theoremstyle{thmstylethree}%
\newtheorem{definition}{Definition}%

\begin{document}

\title[Minimum contrast for the first-order intensity estimation of spatial and spatio-temporal point processes]{Minimum contrast for the first-order intensity estimation of spatial and spatio-temporal point processes}


\author*[1]{\fnm{Nicoletta} \sur{D'Angelo}}\email{nicoletta.dangelo@unipa.it}

\author[2]{\fnm{Giada} \sur{Adelfio}}\email{giada.adelfio@unipa.it}


\affil*[1]{\orgdiv{Department of Economics, Business and Statistics}, \orgname{University of Palermo}, \orgaddress{\city{Palermo}, \country{Italy}}}




\abstract{In this paper, we harness a result in point process theory, specifically the expectation of the weighted $K$-function, where the weighting is done by the true first-order intensity function. This theoretical result can be employed as an estimation method to derive parameter estimates for a particular model assumed for the data. The underlying motivation is to avoid the difficulties associated with dealing with complex likelihoods in point process models and their maximization.
The exploited result makes our method theoretically applicable to any model specification. In this paper, we restrict our study to Poisson models, whose likelihood represents the base for many more complex point process models. 
 In this context, our proposed method can estimate the vector of local parameters that correspond to the points within the analyzed point pattern without introducing any additional complexity compared to the global estimation.
We illustrate the method through simulation studies for both purely spatial and spatio-temporal point processes and show complex scenarios based on the Poisson model through the analysis of two real datasets concerning environmental problems.}




\maketitle

\section{Introduction}

Modeling the complex dynamics of point processes is a fundamental challenge in various fields, ranging from ecology and epidemiology to urban planning. Researchers often deal with the complexities of estimating parameters for specific models due to the nature of their likelihood functions. In this paper, we present a novel approach, based on point process result, that simplifies this task.\\
Typically, a realization of a spatio–temporal point process is characterized via its intensity function, and its parameters are usually fit via the maximum likelihood estimation (MLE) method.\\
Unfortunately, for many point processes, the integral term on the likelihood is often extremely difficult to compute. Even considering the benchmark Poisson process, many choices have to be made for approximating such integral.\\
Approximation methods proposed for certain processes, such as Hawkes processes, suggest a computationally intensive numerical integration method, but in general, the problem of computation or estimation of the integral term in the log-likelihood can be burdensome. Furthermore, Poisson likelihood estimation is proved to be consistent, with its estimates being asymptotically normal, asymptotically unbiased, and efficient, under quite general conditions. However, in real-life applications, we often do not have access to extensive datasets, which can make complex to assess rates of convergence. \\
Despite the computational constraints, Maximum Likelihood (ML) remains the most widely used method for estimating the parameters of point process intensities. The significant work by \cite{baddeley:rubak:tuner:15} has established the Berman-Turner technique \citep{berman1992approximating}  as the predominant approach for fitting parametric Poisson spatial point process models. Furthermore, its spatio-temporal extension continues to be a convenient way to estimate parameters for spatio-temporal point process models, directly building upon purely spatial methodologies.

The scientific literature now recognizes spatial point pattern statistics as a mature discipline, while the spatio-temporal context requires further advancements. Nevertheless, parametric models that depend on external variables present specific challenges. One such challenge is defining the locations of dummy points within the quadrature scheme while respecting the locations of the covariates or, conversely, ensuring knowledge of the covariate values at the selected dummy point locations.

Due to these considerations, while our method is theoretically applicable to various model specifications, this paper confines its scope to Poisson models as a starting proposal for future development. It is important to note that Poisson models cover a diverse range of models, all of which are based on maximizing the Poisson likelihood. The examples covered in this manuscript include homogeneous processes, inhomogeneous processes dependent on both spatial and temporal coordinates, inhomogeneous processes influenced by external covariates, models with spatial and spatio-temporal varying parameters, as well as the estimation of the first-order intensity function in a log-Gaussian Cox process.

In this paper, a novel estimator for the parameters governing spatio–temporal point processes is proposed. Unlike the ML estimator, the proposed estimator does not require the computation or approximation of the computationally expensive integral,  as typically found in the point process log-likelihood, making it computationally more efficient.

The proposed parametric estimator is based on the $K$-function \citep{ripley:76,gabriel2009second} and its deviation from the theoretical value.
This technique, either based on the $K$-function of the pair correlation function, is commonly referred to as the \textit{minimum contrast} technique. Traditionally, it serves as a convenient model-fitting procedure for estimating second-order parameters in a class of inhomogeneous spatial point processes. However, in this context, we utilize it to estimate the parameters of a first-order intensity function.
\\
Our method builds upon a key result in point process theory: the expectation of the weighted $K$-function based on the true first-order intensity function regardless of the parametric form assumed for the model. 
The major intuition of our idea relies upon the fact that such $K$- function weighted by the true first-order intensity function does not identify a specific model among competitor ones, but, conditionally on the parametric form assumed for the data, it is able to identify the best set of parameters of the specified model.
In other words, the weighted $K$-function is now used not only to diagnose a set of competing models and consequently to select the best one but also to select the best model among competitor ones with the same parametric specification. 
\\
A further notable advantage of our method lies in its ability to exploit local second-order characteristics \citep{adelfio2020some}.
Indeed, a model with constant parameters may not adequately represent detailed local variations in the
data, since the pattern may present spatial and temporal variations due to the influence of covariates, the
scale or spacing between points, and also perhaps due to the abundance of points \citep{d2022local}. Indeed, a different
way of analysing a point pattern can be based on local techniques identifying specific and undiscovered
local structure, for instance, sub-regions characterized by different interactions among points, intensity and
influence of covariates \citep{d2023locally}.
By considering the local version of the weighted $K$-function \citep{d2023local}, our approach accurately estimates the vector of local parameters corresponding to specific points within the analyzed point pattern. This level of detail in estimation is crucial for understanding the different variations within spatial and spatio-temporal point processes.\\
Throughout this paper, we will demonstrate the methodology within the spatio-temporal context. It's important to note that every aspect presented can be straightforwardly reduced to the purely spatial context, as illustrated in both the numerical experiments and the applications to real data.\\
All the analyses are carried out through the statistical software \cite{R}.
Section \ref{sec:preliminaries} sets the preliminaries of spatio-temporal point processes, their first- and second-order characteristics, and recalls the most used method employed in literature for fitting \textit{global} and \textit{local} Poisson processes. Section \ref{sec:preliminaries} introduces the new idea for fitting a general Poisson process model through the minimum contrast procedure (MC) and formalizes the estimation procedure.
Section \ref{sec:simulations} presents numerical experiments to study and assess the performance of the proposed fitting procedure, and Section \ref{sec:complex} shows more complex challenges through applications to real datasets.
The paper ends with some conclusions in Section \ref{sec:conclusions}.

\section{Preliminaries
}\label{sec:preliminaries}

Let us give a simple point process defined in space and time, which is a random countable subset $X$ of $\mathbb{R}^2 \times \mathbb{R}$.
Every point $({u}, t) \in X$ corresponds to an event that occurred at a spatial location $ {u} \in \mathbb{R}^2$ and at time $t \in \mathbb{R}$.
Its realization is a finite set $\{({u}_i, t_i)\}^n_{
i=1}$ of distinct points, $n \geq 0$ not fixed in
advance.  This spatio-temporal realization is assumed to occur within a
bounded region $W \times T \subset \mathbb{R}^2 \times \mathbb{R}$, with area and length $|W| > 0, |T| > 0$.

Any event close in both space and time to a given one $({u}, t)$ can be defined by a spatio-temporal cylindrical neighbourhood of the event for each spatial distance $r$ and time lag $h$.
This can be expressed by the Cartesian product
$
b(({u}, t), r, h) = \{({v}, s) : \vert\vert{u} - {v}\vert\vert \leq r, \vert t - s\vert \leq h\} , 
({u}, t), ({v}, s) \in W \times T,
$
with  $\vert\vert \cdot \vert\vert $ denoting the Euclidean distance in $\mathbb{R}^2$.
This will be a cylinder with centre ({u}, t), radius $r$, and height $2h$.

The Campbell Theorem states that, for any non-negative function $f$ on $( \mathbb{R}^2 \times \mathbb{R} )^k$, the following holds
\begin{equation*}
  \mathbb{E} \Bigg[ \sum_{\zeta_1,\dots,\zeta_k \in X}^{\ne} f( \zeta_1,\dots,\zeta_k)\Bigg]=\int_{\mathbb{R}^2 \times \mathbb{R}} \dots \int_{\mathbb{R}^2 \times \mathbb{R}} f(\zeta_1,\dots,\zeta_k) \lambda^{(k)} (\zeta_1,\dots,\zeta_k) \prod_{i=1}^{k}\text{d}\zeta_i.
\label{eq:campbell0}  
\end{equation*}
This essential result defines one of the main tools of point process theory, i.e. the \textit{product densities} $\lambda^{(k)}$, $k  \in \mathbb{N} \text{ and }  k  \geq 1 $.

The arguably most important product densities are obtained for $k=1$ and $k=2$, called the \textit{intensity function} $\lambda$ and the \textit{(second-order) product density} $\lambda^{(2)}$, respectively.

In short, the intensity function gives the rate of occurrence of events in the given region, and the second-order product density describes the correlation between pairs of points of the pattern.

The pair correlation function
\begin{equation*}
    g(({u},t),({v},s))=\frac{ \lambda^{(2)}(({u},t),({v},s))}{\lambda({u},t)\lambda({v},s)}
\end{equation*}
is linked to $\lambda^{(2)}$, formally interpretable as the standardized probability density of an event occurring in two small spatio-temporal volumes. This constitutes an important second-order tool, knowing that a Poisson process has $g(({u},t),({v},s))=1$.

\subsection{
Likelihood-based inference for spatio-temporal Poisson point processes
}

Assuming that the template model is a Poisson process, with a parametric intensity or rate function $\lambda({u}, t; \boldsymbol{\theta}),   u \in
W,  t \in T$, with parameters  $\boldsymbol{\theta} \in \Theta,$ the log-likelihood is 
\begin{equation}
    \log L(\boldsymbol{\theta}) = \sum_i
\lambda({u}_i, t_i; \boldsymbol{\theta}) - \int_W\int_T
\lambda({u}, t; \boldsymbol{\theta}) \text{d}t\text{d}u.
\label{eq:glo_lik}
\end{equation}
In practice, intensity models of log-linear form
$   \lambda({u}, t; \boldsymbol{\theta}) = \exp(\boldsymbol{\theta} Z({u}, t))
$,
are often considered, with $Z({u}, t)$ a spatio-temporal covariate function, including the space or time coordinates themselves. \\
The most direct approach to fitting this model is to adopt the method described by \cite{berman1992approximating}, which involves employing a finite quadrature approximation for the log-likelihood. It is actually the default implemented in the \texttt{spatstat} package \citep{spat}, and for this reason, this approach is widely recognized as the standard method for fitting Poisson spatial models.\\
Renaming the data points as ${x}_1,\dots , {x}_n$ with $({u}_i,t_i) = {x}_i$ for $i = 1, \dots , n$, then $m$  additional \textit{dummy points} $({u}_{n+1},t_{n+1}) \dots , ({u}_{m+n},t_{m+n})$ are generated, to
form a set of $n + m$ quadrature points, where $m$ is commonly taken larger than $n$. Then, some \textit{quadrature weights} $a_1, \dots , a_m$
are determined so that integrals in equation \eqref{eq:glo_lik} can be approximated by a Riemann sum
$    \int_W \int_T \lambda({u},t;\boldsymbol{\theta})\text{d}t\text{d}u \approx \sum_{k = 1}^{n + m}a_k\lambda({u}_{k},t_{k};\boldsymbol{\theta}).
$
The quadrature weights  $a_k$ are taken such that $\sum_{k = 1}^{n + m}a_k = l(W \times T)$, with $l$ the Lebesgue measure.
Then the log-likelihood in equation \eqref{eq:glo_lik} of the template model can be approximated as
\begin{equation}
    \log L(\boldsymbol{\theta}) \approx
\sum_k
a_k
(y_k \log \lambda({u}_k, t_k; \boldsymbol{\theta}) - \lambda({u}_k, t_k; \boldsymbol{\theta}))
+
\sum_k
a_k,
\label{eq:approx}
\end{equation}
taking $y_k = e_k/a_k$, with the indicator $e_k$ equaling $1$ if $u_k$ is a data point and $0$ otherwise.
Apart from the constant $\sum_k a_k$, this expression is formally equivalent to the weighted log-likelihood of
a Poisson regression model. 
This connection to the log-likelihood of a Poisson regression model makes the use of standard Generalized Linear Models (GLM) software possible to maximize it, which significantly contributes to its widespread popularity.    
However, many choices have to be made in order to define the spatio-temporal quadrature scheme. The first one regards the spatio-temporal partition of $W \times T$ into cubes $C_k$ of equal volume $\nu$, and assigning the weight $a_k=\nu/n_k$  to each quadrature point (dummy or data) where $n_k$ is the number of points that lie in the same cube as the point $u_k$. \\
The number of dummy points should be sufficient for an accurate estimate of the likelihood, but at the moment of writing, there aren't guidelines on this aspect. Only \cite{raeisi2021spatio} and \cite{d2023locally}, studying more complex models than the Poisson one,  start with a number of dummy points $m \approx 4 n$, increasing it until $\sum_k a_k = l(W \times T)$.\\
Sometimes, however, a model with constant parameters may not adequately represent detailed local variations in the
data \citep{d2022local}. A local estimation approach should accurately estimate the vector of local parameters corresponding to specific points within the analyzed point pattern \citep{d2023locally}. This level of detail in estimation can reveal to be crucial for understanding the observed variations within space and time.\\
Assume now that the template model is a Poisson process, with a parametric
intensity or rate function $\lambda({u}, t; \boldsymbol{\theta}_i)$  with space
and time locations ${u} \in W,  t \in T$ and parameters $\boldsymbol{\theta}_i \in \Theta$, with $i$ the data point index. 
Estimation can still be performed through the fitting of a GLM using a localized version of the quadrature scheme just introduced.\\
In order to obtain the local estimates $\hat{\boldsymbol{\theta}}_i$, the  local log-likelihood associated with the  spatio-temporal location $({v},s)$ can be written as 
$$
      \log L(({v},s);\boldsymbol{\theta}) = \sum_i w_i(u_i-v,t_i-s)
\lambda({u}_i, t_i; \boldsymbol{\theta})  - \int_W \int_T
\lambda({u}, t; \boldsymbol{\theta}) w_i(u_i-v,t_i-s)\text{d}t \text{d}u  
$$
where for instance $w_i({v},s) = w_{\sigma_s}({v} - {u}_i) w_{\sigma_t}(s - t_i)$. In this case, $w_{\sigma_s} $ and $w_{\sigma_t}$ are weight functions, and $\sigma_s, \sigma_t > 0$ are their smoothing bandwidths. It is not
necessary to assume that $w_{\sigma_s}$ and $w_{\sigma_t}$ are probability densities. For simplicity, one might consider only kernels of fixed
bandwidth, even though spatially adaptive kernels could also be used.
Note that if the template model is the homogeneous Poisson process with intensity $\lambda$, then the local
likelihood estimate of $\hat{\lambda}({v}, s)$ reduces to the kernel estimator of the point process intensity \citep{diggle2013statistical} with
kernel proportional to $w_{\sigma_s}w_{\sigma_t}$.\\
A similar approximation of that used in equation \eqref{eq:approx} for the local log-likelihood associated
with each desired location $({v},s) \in W \times T$ can therefore be used as follows
\begin{equation}\label{eq:quad_loc}
    \log L(({v},s); \boldsymbol{\theta}) \approx
\sum_k
w_k({v},s)a_k
(y_k \log \lambda({u}_k,t_k; \boldsymbol{\theta}) - \lambda({u}_k,t_k; \boldsymbol{\theta}))
+
\sum_k
w_k({v},s)a_k.
\end{equation} 
We refer to \cite{d2023locally} for further details, but basically, for each
desired location $({v},s)$, one replaces the vector of quadrature weights $a_k$ by $a_k({v},s)= w_k({v},s)a_k$, and consequently can still use the GLM software to fit the Poisson local regression.

\subsection{The spatio-temporal $K$-function and its estimator}

\cite{gabriel2009second} define the spatio-temporal inhomogeneous $K$-function and propose a non-parametric estimator.

\begin{definition}{} 
\textit{A   point process defined in space and time is second-order intensity reweighted stationary and isotropic if its intensity function is bounded away from zero and its pair correlation function depends only on the spatio-temporal difference vector $(r,h)$, where $r=||{u}-\textbf{v}||$ and $h=|t-s|$.}
\end{definition}

\begin{definition}{} 
\textit{For a second-order intensity reweighted stationary, isotropic spatio-temporal point process, the spatio-temporal inhomogeneous $K$-function is} 
\begin{equation*}
    K(r,h)=2 \pi \int_{0}^{r} \int_0^{h} g(r',h')r'\text{d}r'\text{d}h'
\end{equation*}
\textit{where $g(r,h)=\lambda^{(2)}(r,h)/(\lambda({u},t)\lambda({v},s)), r=||{u}-{v}||,h=|t-s|$.}
\end{definition}
\textit{The most widely used and simplest estimator of the spatio-temporal $K$-function is:}
\begin{equation}
    \hat{K}(r,h)=\frac{1}{|W||T|}\sum_{i=1}^n \sum_{j > i}^n \mathbf{1}(||{u}_i-{u}_j||\leq r,|t_i-t_j| \leq h).
    \label{eq:k}
\end{equation}

A homogeneous Poisson process has $\mathbb{E}[\hat{K}(r,h)]=\pi r^2 h$, regardless the first-order intensity $\lambda$.
The spatio-temporal $K$-function represents a useful tool to measure interaction and clustering in space and time.
The estimator $\hat{K}(r,h)$ is commonly compared to its theoretical counterpart $\mathbb{E}[\hat{K}(r,h)]=\pi r^2 h$. Values $\hat{K}(r,h) >  \pi r^{2} h$ suggest spatio-temporal clustering of points, while $\hat{K}(r,h) < \pi r^2 h$ suggests a regular pattern.

The inhomogeneous version of the $K$-function in equation \eqref{eq:k} \citep{gabriel2009second} is 
\begin{equation}
        \hat{K}_I(r,h)=\frac{|W||T|}{n(n-1)}\sum_{i=1}^n \sum_{j > i} \frac{\mathbf{1}(||{u}_i-{u}_j||\leq r,|t_i-t_j| \leq h)}{\hat{\lambda}({u}_i,t_i)\hat{\lambda}({u}_j,t_j)}.
    \label{eq:kinh}
\end{equation}
Also, for the inhomogeneous case $\mathbb{E}[\hat{K}_I(r,h)]=\pi r^2 h$,  when the weighting intensity is the true one.
This represents a very important result in the spatio-temporal point process theory since it allows the usage of the weighted estimator $\hat{K}_I(r,h)$ as a diagnostic tool for a general fitted first-order intensity function $\lambda(\cdot,\cdot)$. In other words, it can be used for assessing the goodness-of-fit of spatio-temporal point processes with any fitted first-order intensity function.
In practice, if the fitted intensity is close enough to the true one, its expectation should be close to the theoretical Poisson one $\mathbb{E}[\hat{K}(r,h)]=\pi r^2 h$. Therefore, values $\hat{K}_I(r,h)$ greater than $\pi r^{2} h$ indicate that the model is not a good fit since the distances among points exceed those of the theoretical  Poisson.
 
In \cite{adelfio2020some}, local versions of both the homogeneous and inhomogeneous spatio-temporal $K$-functions are provided,  as diagnostic tools accounting also for local characteristics.
They define an estimator of the intensity by $\hat{\lambda}=n/(|W||T|)$, and then propose expressing the localized version of equation \eqref{eq:k}  for the $i$-th event $({u}_i,t_i)$ as  
\begin{equation}
    \hat{K}^i(r,h)=\frac{1}{\hat{\lambda}^2|W||T|}\sum_{(\textbf{u}_i,t_i)\ne ({v},s)} \mathbf{1}(||{u}_i-{v}||\leq r,|t_i-s| \leq h)
    \label{eq:kl}
\end{equation}
and the local version of equation \eqref{eq:kinh} as
\begin{equation}
    \hat{K}^i_{I}(r,h)=\frac{1}{|W||T|}\sum_{({u}_i,t_i)\ne ({v},s)} \frac{\mathbf{1}(||{u}_i-{v}||\leq r,|t_i-s| \leq h)}{\hat{\lambda}({u}_i,t_i)\hat{\lambda}({v},s)},
    \label{eq:kinhl}
\end{equation}
with $({v},s)$ being any other point's spatial and temporal coordinates.
  They further proved that also the local inhomogeneous estimators behave as the corresponding theoretical Poisson ones, i.e. the expectation of equations \eqref{eq:kl} and \eqref{eq:kinhl} is  $\pi r^2 h$ as well.

\section{Minimum contrast for first-order intensity estimation}\label{sec:mc}

\begin{figure}[H]
	\centering
	\includegraphics[trim={0 0 0 3cm},clip,width=.5\textwidth]{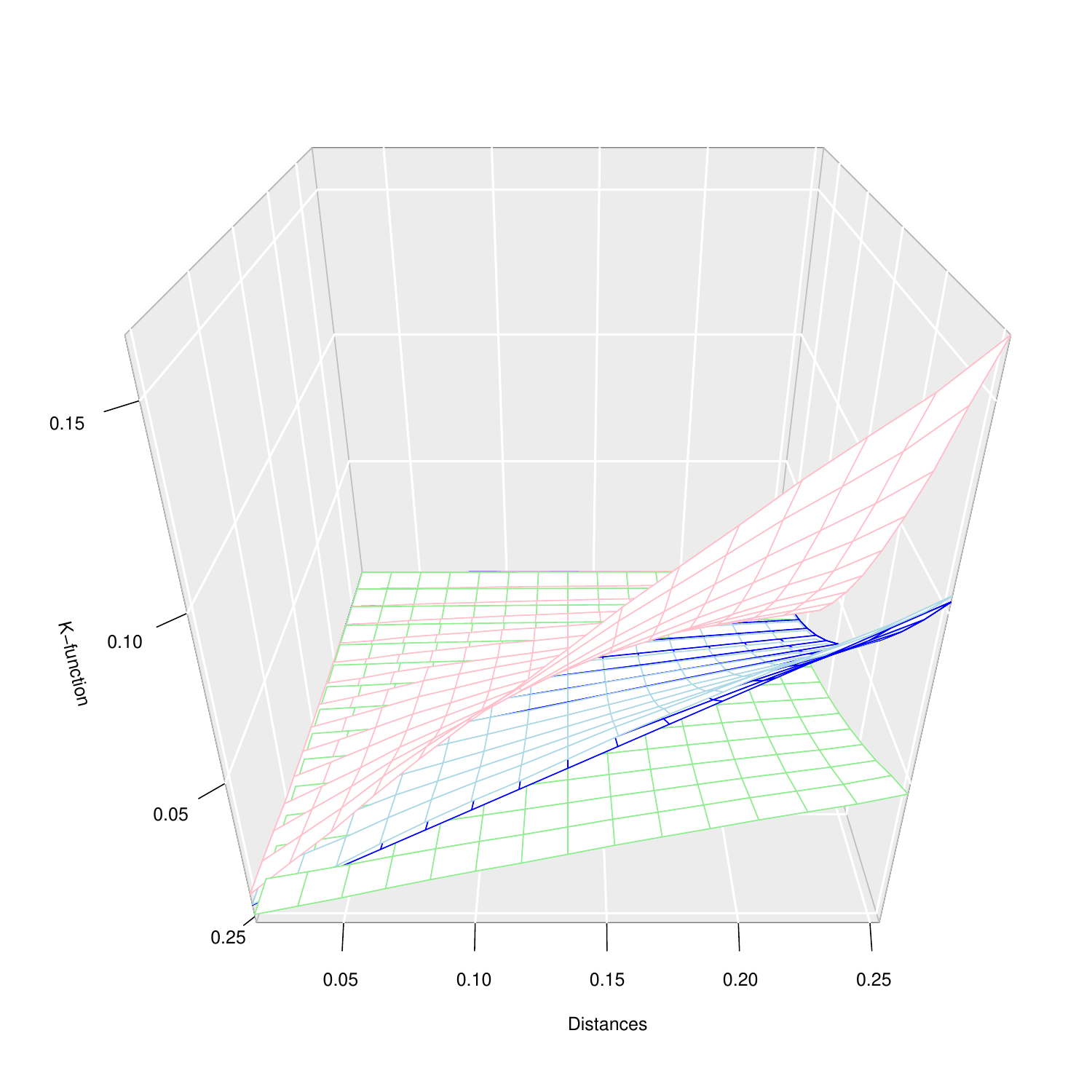}

	\caption{In blue: the theoretical $K$-function of a simulated Poisson process with 500 points. In light blue: the estimated $K$-function, weighted by the true intensity function. In pink and green: the estimated $K$-functions, weighted by some wrong intensities ($\lambda = \{400, 740 \}$).}
	\label{fig:motivation}
\end{figure}

In this section, we present the rationale behind our intuition for employing the $K$-function as an inferential tool for estimating model parameters in a minimum contrast approach. In particular, we refer to a graphical representation of a straightforward example, the homogeneous Poisson process model, characterized by a constant intensity.\\
Our primary insight arises from the realization that the $K$-function, when weighted by the true first-order intensity function, can serve as a tool for identifying the optimal set of parameters for the assumed parametric model, as opposed to the traditional approach of selecting the best model among competing alternatives.\\
This idea is shown graphically in Figure \ref{fig:motivation}.

The blue surface of Figure \ref{fig:motivation} represents the theoretical $K$-function of a simulated spatio-temporal Poisson process with 500 points. As an example, we estimate three $K$-functions, respectively weighted by: the true intensity (in light blue) and two wrong constant intensities (400 in pink and 750 in green).
From these plots, we can observe that the overall behaviour of the $K$-function is the same (i.e. increasing with the space and time lags), with the only difference in the magnitude values of the $K$-functions. In particular, the $K$-function weighted by the true intensity function reports the values closer to the theoretical one.\\
Indeed, the theory suggests that the squared difference between the observed $K$-function and the theoretical one should approach zero, as the intensity used for weighting the observed $K$-function approaches the true one \citep{adelfio2020some}.
The value of the sum of the squared differences are $0.07$, $0.0002$ and $0.06$ for the $K$-functions weighted by $400$, $500$ and $750$, respectively. As expected, the lowest value is obtained when weighting for the true intensity function.

We now formalize the proposed method of parameter estimation using $K_I(r,h)$ in equation \eqref{eq:k} and its estimator $\hat{K}_I(r,h)$ of equation \eqref{eq:kinh}. 

Let be given a point process model with the intensity $\lambda({u},t;\boldsymbol{\theta})$ (in brief: $\lambda_{\boldsymbol{\theta}}$) with a vector of parameters $\boldsymbol{\theta} \in \Theta$.
The proposed \textit{minimum contrast for first-order intensity estimation} (MC) procedure is defined by the minimization of the following objective function
 \begin{equation}
\mathcal{M}(\boldsymbol{\theta})= \int_{h_0}^{h_{max}}\int_{r_0}^{r_{max}}\phi(r,h)\{(\hat{K}_I(r,h; \lambda_{\boldsymbol{\theta}})) - \pi r^2 h) \}^2 \text{d}r \text{d}h
    \label{eq:min_con}
\end{equation}
with respect to $\boldsymbol{\theta}$, providing a vector of estimates $\hat{\boldsymbol{\theta}}$:

\begin{equation*}
\hat{\boldsymbol{\theta}} = \text{arg} \min_{\boldsymbol{\theta} \in \Theta} \mathcal{M}(\boldsymbol{\theta}).
\label{eq:min_con_est}
\end{equation*}

Here  $r_0$, $h_0$, $r_{max}$ and $h_{max}$ are the lower and upper space and time lag limits of the contrast criterion, and $\phi(r, h)$ is a weight that depends on the spatio-temporal distance.
Note that our MC proposal poses the basis on the assumption that $\pi r^2 h$ is the expected value of the $K$-function when this is weighted by the true intensity function. Given the model assumed for the data, the combination of parameters leading to a good intensity (that is, a small discrepancy in equation \eqref{eq:min_con}) may not be unique. This might result in biased (and unreliable) estimates. We refer to  \cite{baddeley2022fundamental} for a detailed classification of causes for these practical difficulties, with particular reference to the purely spatial case.

For this reason, we employ a further step in our proposal, by a radial penalization presented in \cite{kreutz2018easy}.
That work aimed at addressing the feasibility of unique parameter estimation in dynamic systems.
Numerical optimization is used to test the uniqueness of parameters by means of a penalty in the radial direction which has the goal to enforce the displacement of the parameters.
The suggested method is based on a comparison of the objective function of common fitting with a penalized fit pulling the parameter vector away from the first estimate. A major characteristic of their method is that it allows identifiability
analysis of all parameters in a joint manner, as well as the possibility to investigate the identifiability of each parameter individually. Their proposed approach enables quick testing for parameter identifiability, whereas the several other approaches proposed in the literature are typically computationally demanding, difficult to perform and/or not applicable in many application settings.
Indeed, their method is suited for any model with an optimization-based parameter estimation such as \textit{maximum likelihood} and \textit{least-squares}. This is the reason why it appears applicable to our purposes, particularly addressing the joint identifiability investigation of all model parameters.

\begin{definition}
\textit{Let $\boldsymbol{\theta} \in \Theta$ be the parameter vector containing all $p$ unknown constants in the model assumed for the data.
A parameter $\theta_j$ ($j = 1, \ldots, p$) is said to be structurally locally identifiable, if, for almost any $\theta_{j}$, there exists a neighbourhood $P$ such that if}
\begin{equation*}
\boldsymbol{\theta} \in P \quad \land \quad g(\theta_{j}^{(1)}) = g(\theta_{j}^{(2)})
 \quad   \Rightarrow   \quad \theta_{j}^{(1)} = \theta_{j}^{(2)} 
\end{equation*}
\textit{``Almost any'' means for all parameters except for isolated points. If this property holds not only within a neighbourhood but also for the whole parameter space, the parameter is termed structurally globally identifiable} \citep{chis2011structural}. 
\end{definition}

Therefore, alternatively to estimating $\boldsymbol{\theta}$ according to equation \eqref{eq:min_con}, we suggest the use of a penalized objective function 
\begin{equation}
    \mathcal{M}_{tot}^{R}(\boldsymbol{\theta}) =  \mathcal{M}(\boldsymbol{\theta}) +  \mathcal{M}_{pen}^{R}(\boldsymbol{\theta})
    \label{eq:tot}
\end{equation}
with 
\begin{equation}
 \mathcal{M}_{pen}^{R}(\boldsymbol{\theta}) = 
 \frac{1}{R^2}
 \Biggl(
 \sqrt{\sum_{j}(\theta_j - \hat{\theta}_j)^2}
 -R\Biggl)^2,
 \label{eq:pen}
\end{equation}
where
$1/R^2$ is the tuning parameter representing the penalization strength. 
In essence, the penalty acts as an extra data point that is utilized to pull the parameter in the direction where the data is least informative.

The penalty term $ \mathcal{M}_{pen}^{R}(\boldsymbol{\theta})$ has its minimum at a sphere with radius $R$ centered around $\hat{\boldsymbol{\theta}}$. The final estimates are found as 
\begin{equation*}
{\boldsymbol{\theta}}^{*} = \text{arg} \min_{\boldsymbol{\theta} \in \Theta} \mathcal{M}^{R}_{tot}(\boldsymbol{\theta}).
\label{eq:min_con_tot_est}
\end{equation*}

\subsection{Selection of the radius R}\label{sec:R}

Any penalized approach requires a tuning parameter which controls the degree of shrinking of the model coefficients.
Traditional selection criteria for regression problems involve selecting the lowest goodness-of-fit criterion among competitors.
However, the main problem when dealing with residual analysis for point processes is to find a correct definition of residuals since the one used in dependence models can not be used for point processes \citep{adelfio2020some}.\\
An alternative approach to defining a weighted second-order statistic is the usage of the smoothed raw residuals \citep{baddeley2005residual}.

As our whole fitting method is based on a weighted second-order statistic, we choose to employ the smoothed raw residuals for the selection of the radius $R$.

The predicted number of points occurring in a given spatial region $W$ is equal to $\int_W \hat{\lambda}_{\boldsymbol{\theta}}({u})$d${u}$, with $\hat{\lambda}_{\boldsymbol{\theta}}({u})$ the intensity of a model fitted to an inhomogeneous Poisson process. Consequently, the \textit{raw residual process} \citep{baddeley2005residual} in each region $W \subset \mathbb{R}^2$ can be defined as the number of points which fall in $W$, 
denoted as 
\begin{equation}
    r(W)=n(\textbf{x} \cap W) - \int_W \hat{\lambda}_{\boldsymbol{\theta}}({u})\text{d}{u}.
    \label{eq:raw}
\end{equation} Here \textbf{x} denotes the observed realization of a purely spatial point pattern, and $n(\textbf{x} \cap W)$ is its number of points falling in  $W$.
Increments of $r(W)$ are analogous to the raw residuals (observed minus fitted values) in a linear model. The adequacy of the fitted model can be checked by inspecting whether $r(W) \approx 0$ and various plots and transformations of $r(W)$ can be useful diagnostics for a fitted point process model.
The resulting residuals can be displayed easily by smoothing them.
Hence, we can define the \textit{smoothed residual fields} as
\begin{equation}
s({u})=\tilde{\lambda}_{\boldsymbol{\theta}}({u})-\lambda^{\dagger}({u})
	\label{eq:smo0}
\end{equation}
with $\tilde{\lambda}_{\boldsymbol{\theta}}({u}) 
$  a non-parametric estimate of the fitted intensity, usually estimated through kernel procedures. 
A common practice is to select the smoothing bandwidth for the kernel estimation of the raw residuals by cross-validation as the value that minimizes the Mean Squared Error criterion defined by \cite{diggle1985kernel}, by the method of \cite{berman1989estimating}. This is because, among the possible alternatives proposed in the literature, the cross-validation method is typically the most adaptive, and, therefore, the one which should resemble the true intensity function the most. This is particularly relevant when assessing the goodness-of-fit of a parametric fitted intensity function. See \cite{diggle2013statistical} for further details.
On the other hand, $\lambda^{\dagger}({u})$ is a smoothed version of the estimated intensity of the fitted model. 
Note that the smoothed residual fields procedure is intended for parametric specifications of the fitted intensity function. 
Given these, of course, smaller differences in equation \eqref{eq:smo0} indicate that the fitted model is close to the real one.
For this reason, we choose the best model among competitors as the one with the lowest values of the smoothed raw residuals.

A note is in order. Raw residuals in equation \eqref{eq:raw} are proposed by \cite{baddeley2000non}. 
Whereas previous works \citep{lawson1993deviance,stoyan1991second} have defined diagnostic values for the data points only, these residuals are also ascribed
to locations which are not points of the pattern. This is related to an important methodological issue for point processes. In a point pattern dataset, the observed information isn't limited to the locations of the observed points within the pattern. The absence of points in other locations also provides valuable information.
With the additional advantage of graphical presentation, smoothed raw residuals become a straightforward and effective diagnostic tool.

We, therefore, select the radius $R$ as the one minimizing the integration of the smoothed raw residuals over the analysed area:
\begin{equation}
\hat{R} = \text{arg} \min_{R} \int_W (\tilde{\lambda}({u})-\lambda^{\dagger}_R({u})) \text{d}{u},
\label{eq:chooseR}
\end{equation}
where $\lambda^{\dagger}_R({u}) = \lambda^{\dagger}({u}; \hat{\boldsymbol{\theta}}_R)$, which is the intensity obtained by imputing the estimated parameter vector $\hat{\boldsymbol{\theta}}$ with the $R$ radius value in the penalization procedure.

Moving to the spatio-temporal context,  fixed bandwidths for spatial or spatio-temporal data based on the maximal smoothing (over-smoothing) principle of \cite{terrell1990maximal} can be employed. The optimal values minimize the asymptotic mean integrated squared error assuming normally distributed data \citep{silverman:86}. Two separate values can be obtained for the purely spatial and temporal components by independently applying the normal scale rule to the spatial and temporal margins of
the supplied data.
Alternatively, bandwidth selection for standalone spatio-temporal density/intensity can be based on either unbiased least
squares cross-validation (LSCV), likelihood (LIK) cross-validation, or bootstrap estimation
of the MISE, providing an isotropic scalar
spatial bandwidth and a scalar temporal bandwidth.

\subsection{Local extension}\label{sec:proposal}

Our proposal can also be extended in a local context, representing an alternative to the most standard local likelihood procedure, which consists of many choices, including the quadrature scheme but also the weighting functions for the spatio-temporal kernels (as in equation \eqref{eq:quad_loc}).
Suppose then that the model incorporates a vector of parameters $\boldsymbol{\theta}_i$ for each point $i$. Let $\hat{K}_I^i(r,h; \lambda_{\boldsymbol{\theta}}))$ denote the local estimators calculated from the data.

For each point indexed by $i$ we consider
\begin{equation*}
\mathcal{M}_{local}(\boldsymbol{\theta}_i) =  \int_{h_0}^{h_{max}}\int_{r_0}^{r_{max}}\phi(r,h)\{(\hat{K}_I^i(r,h; \lambda_{\boldsymbol{\theta}}) - \pi r^2 h) \}^2 \text{d}r \text{d}h.
\end{equation*}

Then, we can obtain a vector of estimates $\hat{\boldsymbol{\theta}}_i$, one for each point $i$, as 
\begin{equation*}
\hat{\boldsymbol{\theta}}_i = \text{arg} \min_{\boldsymbol{\theta}\in \Theta} \mathcal{M}_{local}(\boldsymbol{\theta}_i).
\end{equation*}

In practice, one could also wish to specify a different weight function $\phi(r,h)$ for each point $i$, making the objective function as follows 
\begin{equation*}
\mathcal{M}_{local}(\boldsymbol{\theta}_i) = \int_{h_0}^{h_{max}}\int_{r_0}^{r_{max}}\phi_i(r,h)\{(\hat{K}_I^i(r,h; \lambda_{\boldsymbol{\theta}}) - \pi r^2 h) \}^2 \text{d}r \text{d}h.
\end{equation*}

Finally, the penalized optimization could also be implemented locally, giving rise to the estimated parameters
\begin{equation*}
{\boldsymbol{\theta}}^{*}_i = \text{arg} \min_{\boldsymbol{\theta}\in \Theta} \mathcal{M}^{R}_{local;tot}(\boldsymbol{\theta}_i),
\end{equation*}
where the component $\mathcal{M}_{local;pen}^{R}(\boldsymbol{\theta}_i)$
could depend on a fixed $R$ value, or either on an individual one $R_i$ for each point.

\subsection{Cox processes}\label{sec:cox}

Cox processes are point process models typically used when observing clustering among points of the analysed pattern. 
In general, any Cox model can be estimated by a two-step procedure, involving first the first-order intensity and then the cluster or correlation parameters. \\
First, a Poisson process with a particular model for the log-intensity is fitted to the point pattern data, providing the estimates of the coefficients of all the terms that characterize the intensity. \\
Then, the estimated  intensity is  taken as the true one and the cluster  or correlation parameters are estimated using either the \textit{method of minimum contrast} \citep{pfanzagl1969measurability,eguchi1983second,diggle1979parameter,diggle1984monte,moller1998log,davies2013assessing,siino2018joint}, \textit{Palm likelihood} \citep{ogata1991maximum,tanaka2008parameter}, or
 \textit{composite likelihood} \citep{guan2006composite}.\\
   	The most common technique for this second stage is the \textit{minimum contrast}, and it is the method which we shall refer to here.\\
Log-Gaussian Cox processes are one of the most prominent clustering models.
By specifying the intensity of the process and the moments of the underlying Gaussian Random Field (GRF), it is possible to estimate both the first and second-order characteristics of the process. 
Following the inhomogeneous specification in \cite{diggle2013spatial}, a LGCP for a generic point in space and time has the intensity
	$		\Lambda({u},t)=\lambda({u},t)\exp(S({u},t))
	$
where $S$ is a Gaussian process with $\mathbb{E}(S({u},t))=\mu=-0.5\sigma^2$ and so  $\mathbb{E}(\exp{S({u},t)})=1$ and with variance and covariance matrix $\mathbb{C}(S({u}_i,t_i),S({u}_j,t_j))=\sigma^2 \gamma(r,h)$ under the stationary assumption, with $\gamma(\cdot)$ the correlation function of the GRF, and $r$ and $h$ some spatial and temporal distances. Following \cite{moller1998log}, the first-order product density and the pair correlation function of an LGCP are $\mathbb{E}(\Lambda({u},t))=\lambda({u},t)$ and $g(r,h)=\exp(\sigma^2\gamma(r,h))$, respectively.  \\
Driven by a GRF, controlled in turn by a specified covariance structure, the implementation of the LGCP framework in practice requires a proper estimate of the intensity function. Usually, this is achieved through the maximization of the Poisson likelihood in equation \eqref{eq:glo_lik}, with all the challenges already introduced.
We substitute the fitting of the first-order intensity function with the proposed minimum contrast procedure, making the \textit{two-step minimum contrast estimation procedure} as follows:
\begin{itemize}
	\item The intensity parameters $\boldsymbol{\theta}$ are estimated by minimizing either $\mathcal{M}(\boldsymbol{\theta})$ in equation \eqref{eq:min_con} or $\mathcal{M}_{tot}^R(\boldsymbol{\theta})$ in equation \eqref{eq:tot}, depending on the necessity (instead of the maximising likelihood of the Poisson process with intensity $\lambda(\cdot;\boldsymbol{\theta)}$)
	\item The interaction/covariance parameters $\boldsymbol{\psi}$ are estimated by minimizing the discrepancy between a second-order summary statistics (either the pcf or $K$-function) and its theoretical value under the assumed covariance function. 	
\end{itemize}
In particular, we propose to perform the second step relying on the \textit{joint minimum contrast} \citep{siino2018joint} procedure for obtaining the interaction parameters $\boldsymbol{\psi}$ 
$$
	\mathcal{M}_J( \boldsymbol{\psi})=\int_{h_0}^{h_{max}} \int_{r_0}^{r_{max}} \phi(r,h) \{\hat{J}(r,h)-J(r,h;\boldsymbol{\psi})\}^2 \text{d}r \text{d}h,
$$
 with $\hat{J}(r,h)$ the estimate of the second-order summary statistics and $J(r,h;\boldsymbol{\psi})$ its theoretical value depending on the functional form assumed for the covariance structure. \\
 We would like to emphasize that in this paper, particularly in the simulation studies, we execute the first step using the proposed procedure based on the $K$-function, and the second step through the pair-correlation function.

\section{
Simulation results
}\label{sec:simulations}
In this section, we report some experimental results, for illustrating the proposed estimation procedure, both for the space and spatio-temporal case.
Note indeed that the whole theory introduced in section \ref{sec:mc} is easily implementable for the purely spatial case, as the theory regarding the properties of global and local weighted second-order statistics holds the same. We refer to \cite{adelfio2020some} and references therein.

\subsection{Space}

We simulate 1000 spatial point patterns from three different purely spatial point processes in the unit square with 500 points on average following these scenarios:
\begin{enumerate}
    \item  homogeneous Poisson process with constant intensity $\lambda=\exp(\theta_0)$;
    \item inhomogeneous Poisson process with intensity   $\lambda(x,y)=\exp (\theta_1  x)$;
    \item  inhomogeneous Poisson process with intensity   $\lambda(x,y)=\exp (\theta_0 + \theta_1 x)$.
\end{enumerate}

Table \ref{tab:6_res7} shows the results of the proposed minimum contrast estimation procedure, reporting the means and MSE of the estimates obtained for the 1000 replications.
 The spatial lags in the $K$-function are 153 values ranging between $0$ and $1/4$ of the maximum distance.

\begin{table}[!ht]
\centering
\caption{ Means and MSE of the estimates obtained over 1000 simulations for the purely spatial scenarios.}
\begin{tabular}{lll|ccc}
 \toprule
Process&$\mathbb{E}[n]$&True par&mean&MSE&S.E.\\
 \midrule
Homogeneous Poisson &500&$\theta_0 = 6.21$&
6.21& 0.002
&0.03\\
 \midrule
 Inhomogeneous Poisson & 500&$\theta_1 =  8.34$ & 8.64 &2.81&0.69 \\
  \midrule
 Inhomogeneous Poisson  & 500&$\theta_0 =  2$& 3.37&2.12
&20.06 \\
 & &$\theta_1 = 6$ & 4.03&2.06
&23.71 \\
  \midrule  
  Inhomogeneous Poisson  & 500&$\theta_0 =  2$ &2.01  & 2.83
& 0.79\\
 $R\in [0.25, 10]$ Eq. \eqref{eq:bestR} & &$\theta_1= 6$ & 5.96&2.81
& 1.52\\
  \midrule
    Inhomogeneous Poisson  & 500&$\theta_0 =  2$ & 1.94& 2.89
&0.87  \\
 $R = 2.5$ & &$\theta_1 = 6$ &6.02 &2.89
& 1.52\\
\midrule
  Inhomogeneous Poisson & 500&$\theta_0 =  2$ &1.81 & 3.01
&  0.91\\
 $R\in [0.25, 10]$ Eq. \eqref{eq:chooseR} & &$\theta_1 = 6$ &6.22 &3.04
& 1.69\\
  \bottomrule
\end{tabular}
\label{tab:6_res7}
\end{table}

The procedure performs effectively when the parameter to be estimated is unique. However, challenges arise in the third scenario, likely stemming from issues related to parameter identifiability. Indeed, as anticipated in section \ref{sec:mc}, the minimum of the objective function in equation \eqref{eq:min_con} is not unique, but the combination of the estimated parameters represents the best fit in terms of intensity.
Therefore, we proceed by adding a penalty to the already employed objective function only for the third scenario.

Having simulated from known parameters, we choose the radius $R$ to be used in the penalization procedure as the one minimizing the discrepancy between the true parameters and the estimated ones. Formally:
\begin{equation}
\hat{R} = \text{arg} \min_{R} \sum_{j} \frac{({\theta}_j - \hat{{\theta}}_{j,R})^2}{\hat{{\theta}}_{j,R}}.
\label{eq:bestR}
\end{equation}
We explore different ranges for $R$, omitted for brevity in  Table \ref{tab:6_res7}, and we note that the choice of the $R$ range, in the minimization of equation  \eqref{eq:bestR}, does not seem to be relevant. Indeed, the penalization procedure clearly overcomes the identifiability problem previously encountered. Moreover, we can spot smaller standard errors than the unpenalized version.
Another relevant result is that the mean of all the selected $R$ values over the 1000 simulations, i.e. R= 2.5,  leads to comparable results. We interpret this numerical result as an indication that an optimal value of $R$ should exist, which is related to the combination of the true parameters.

Naturally, in real data applications, one does not have the knowledge of the true parameter values, making it impossible to carry out the minimization as described in equation \eqref{eq:bestR}. 
Therefore, we proceed by employing the tuning parameter selection criterion proposed in section \ref{sec:R}. 
 It is important to note that further investigations could be conducted to assess the performance of the $R$ selection procedure as outlined in section \ref{sec:R}. However, at the time of writing, the testing of the $R$ selection procedure falls outside the scope of this research.

\subsection{Space-time}

Moving to the spatio-temporal context, we simulate 1000 spatio-temporal point patterns in the unit cube with 500 points on average from the following point processes:
\begin{enumerate}
    \item  homogeneous Poisson process with constant intensity $\lambda=\exp(\theta_0)$;
    \item  inhomogeneous Poisson process with intensity   $\lambda(x,y,t)=\exp (\theta_0 + \theta_1  x)$.
\end{enumerate}

The spatial and temporal distances used in the observed weighted $K$-functions are 15 values ranging from 0 to $r_{max}$, equal to 1/4 of the maximum (spatial or temporal) distances.

Table \ref{tab:6_res11} reports the means and MSE of the estimates of the two considered processes, averaged over 1000 simulations.

\begin{table}[!ht]
\centering
\caption{Means and MSE of the estimates obtained over 1000 simulations for the spatio-temporal scenarios.}
\begin{tabular}{lll|ccc}
 \toprule
Process&$\mathbb{E}[n]$&True par&mean&MSE&S.E.\\
  \midrule
Homogeneous Poisson &500&$\theta_0  =6.21$&6.27
&0.006
&2.62\\
  \midrule
 Inhomogeneous Poisson  & 500&$\theta_0 =  2$ &2.85 &  2.38
& 1.19 \\
$R=2.5$ & &$\theta_1 = 6$ & 4.89&2.34
& 1.52\\
  \bottomrule
\end{tabular}
\label{tab:6_res11}
\end{table}

We notice that the mean of the intensity function for the homogeneous scenario appears systematically overestimated. 
For the inhomogeneous point processes, we employ the penalized procedure with radius $R = 2.5$ since, as noticed
in the purely spatial case,  this tends to make the estimated parameters more similar to the true values than in the unpenalized case.

\subsubsection{Local space-time}

Section \ref{sec:proposal} introduced the further advantage of our proposal, showing the possibility of fitting local parameters. Preliminary analyses not reported for brevity showed a systematic overestimation of the local parameters in the homogeneous case.
Moving to the inhomogeneous scenario, we show results over 100 simulated patterns in Table \ref{tab:6_res12}.

\begin{table}[!ht]
\centering
\caption{  Mean and quartiles of the distributions of the local parameters,  averaged over 100 simulated point patterns. MSE values come in parentheses.}
\begin{tabular}{lll|cccc}
 \toprule
Process&$\mathbb{E}[n]$&True par&25\%&50\%(MSE)&mean(MSE)&75\%\\
  \midrule
 Inhomogeneous Poisson  & 500&$\theta_0 =  2$ & -1.28&3.55(2.10) 
& 2.59(2.49)&6.08\\
 & &$\theta_1 = 6$ &0.95 &4.34(2.10)
&5.09(2.34) & 9.50\\
  \midrule
 Inhomogeneous Poisson  & 500&$\theta_0 =  2$ &0.95&2.49(2.51)&3.09(2.22)&4.44\\
$R=2.5$& &$\theta_1 = 6$ & 2.87&5.39(2.45)&6.12(2.96)&9.08\\
  \bottomrule
\end{tabular}
\label{tab:6_res12}
\end{table}
The means and medians are similar to the true values of the parameters, and the variability of the local estimates is smaller when adding the penalty.  
For the considered simulation scenarios we highlight that the fitted intensity, obtained imputing the average of the local parameters, greatly resembles the true one.

\subsection{Convergence study}

Having assessed the performance of the proposed procedure, we now proceed to investigate the impact of varying sample sizes in the simulated patterns to gain a deeper understanding of the convergence behavior.\\
At this scope, Tables \ref{tab:6_res7bis} and \ref{tab:6_res11bis} contain the means, MSE, and standard errors (S.E.) of the estimates obtained over 100 simulations of both homogeneous and inhomogeneous processes, for three sample sizes: $\mathbb{E}[n] = \{ 250, 500, 750\}$.\\
Table \ref{tab:6_res7bis} reports results for the purely spatial scenarios, while Table \ref{tab:6_res11bis} contains those of the spatio-temporal one.
In both cases, we employed the $R=2.5$ penalization for the inhomogeneous processes, as it proved to be the optimal value in the previous simulations.

\begin{table}[!ht]
\centering
\caption{Means, MSE, and S.E. of the estimated obtained over 100 simulations for the purely spatial scenarios.}
\begin{tabular}{lll|cccccc}
 \toprule
&&&&MC&&&MLE&\\
Process&$\mathbb{E}[n]$&True par&mean&MSE&S.E.&mean&MSE&S.E.\\
  \midrule
Homogeneous Poisson &250&$\theta_0 = 5.52$&5.51
& 0.004
&0.03&5.51&0.004&0.06\\
 &500&$\theta_0 = 6.21$&
6.21& 0.002
&0.03&6.22&0.002&0.04\\
&750&$\theta_0 = 6.62$&6.62
& 0.001
&0.03&6.62&0.001&0.04\\
  \midrule 
 Inhomogeneous Poisson  & 250&$\theta_0 =  1.3215$&1.49&0.26
&0.95 &1.31&0.10&0.34\\
 $R = 2.5$  & &$\theta_1 = 6$ & 5.59&0.54
&1.44 &6.00&0.14&0.40\\
& 500&$\theta_0 =  2$& 1.89 &0.13
& 0.87&1.95&0.05&0.24\\
  & &$\theta_1 = 6$ &6.09&0.21
&  1.52&6.05&0.07&0.29\\
 & 750&$\theta_0 = 2.42 $&2.27 &0.11
&0.79&2.41&0.04&0.19\\
  & &$\theta_1 = 6$ & 6.29&0.21
& 1.59&6.02&0.06&0.23\\
  \bottomrule
\end{tabular}
\label{tab:6_res7bis}
\end{table}

\begin{table}[!ht]
\centering
\caption{Means, MSE, and S.E. of the estimated obtained over 100 simulations for the spatio-temporal scenarios.}
\begin{tabular}{lll|ccc|ccc}
 \toprule
&&&&MC&&&MLE&\\
Process&$\mathbb{E}[n]$&True par&mean&MSE&S.E.&mean&MSE&S.E.\\
  \midrule
Homogeneous Poisson &250 &$\theta_0  = 5.52$&5.59
& 0.01&2.33&5.59&1.06&0.06\\
&500 &$\theta_0  = 6.21$&6.27
&  0.006&2.62&6.26&0.13&0.04\\
&750 &$\theta_0  = 6.62$&6.68
& 0.005 &2.79&6.69&0.006&0.04\\
  \midrule
 Inhomogeneous Poisson  & 250&$\theta_0 =  1.3215$ &2.43 &  2.21&0.78&1.84&0.33&0.33\\
$R=2.5$ & &$\theta_1 = 6$ & 4.56& 3.46&0.78& 5.37&0.49&0.39\\
& 500&$\theta_0 =  2$ &2.76 &  0.92 & 0.72&2.53&0.33&0.23\\
 & &$\theta_1 = 6$ &  5.01&1.58& 0.74&5.33&0.50&0.27\\
& 750&$\theta_0 =  2.42$ & 3.18&  0.94& 0.71&2.97&0.33&0.19\\
& &$\theta_1 = 6$ & 5.01&  1.65&0.74&5.32&0.50&0.22\\
  \bottomrule
\end{tabular}
\label{tab:6_res11bis}
\end{table}

We have established that the most favorable scenario for our proposal is the purely spatial homogeneous one, characterized by minimal bias and limited variability in the estimates. The sample size primarily impacts the MSE, which naturally decreases with the increasing number of points in the simulated pattern, as expected. \\
In contrast, the spatio-temporal estimates of the homogeneous intensity exhibit a higher level of bias,  providing a slight overestimation of the single parameter.\\
Focusing to the inhomogeneous case, we observe that both the spatial and spatio-temporal scenarios display some degree of bias. Among these scenarios, the one with 500 points stands out as the most favorable.
This is likely due to the value of  $R$ set to $2.5$, which was actually obtained from simulations with that sample size.\\
However, the MSE value shows minimal improvement as the number of points increases, moving from 500 to 750 points, whether in spatial or spatio-temporal contexts.\\
Moving to the comparison with MLE results,  we can note there are differences in the standard errors and performance of these methods in different scenarios. In the spatial homogeneous case, MLE and MC results are pretty similar  and
MLE standard errors are slightly higher than the MC ones. In the homogeneous spatio-temporal case, MLE standard errors are quite smaller than MC standard errors, that is, MLE provides more precise and reliable parameter estimates compared to MC in this particular scenario. In inhomogeneous cases (both spatial and spatio-temporal), we can still note a difference in standard errors between MLE and MC  but of lower magnitude, suggesting that MLE still outperforms MC in terms of parameter estimation, but the difference is not as pronounced as in the homogeneous spatio-temporal case.
Finally, the better performance in spatial contexts is attributed to the additional complexity of the temporal component in the spatio-temporal scenarios.
In summary, as expected, it seems that MLE performs slightly better, mostly for scenarios involving spatio-temporal complexity, as it provides more precise parameter estimates with smaller standard errors compared to MC. However, our proposal may represent a promising and valid alternative, especially when considering the computational complexity of the specific statistical methods and observed data.

\subsection{External covariates}

\begin{figure}[H]
    \centering
\includegraphics[trim={0 0 0 1cm},clip,width=.45\textwidth]{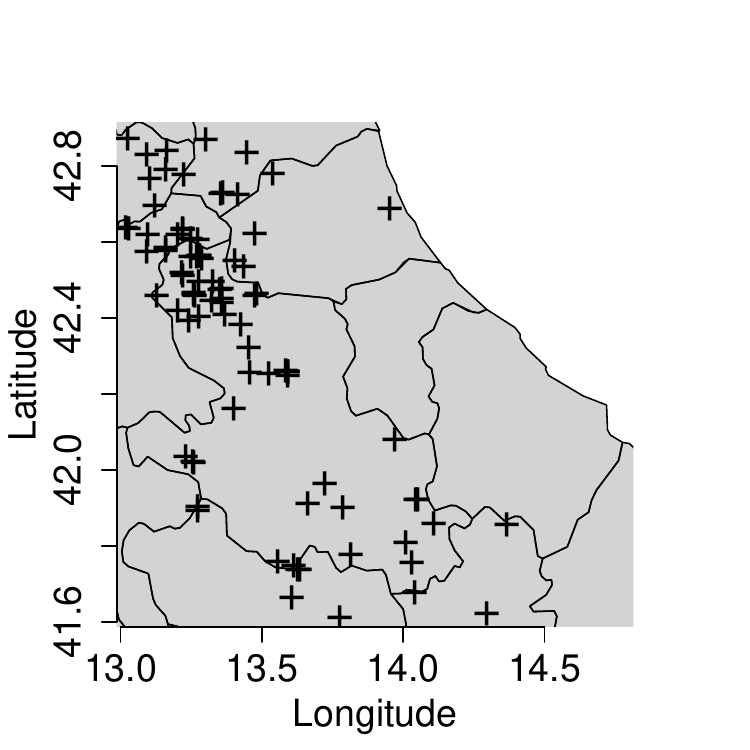}
\includegraphics[trim={1.25cm 2.5cm .5cm 3.25cm},clip,width=.525\textwidth]{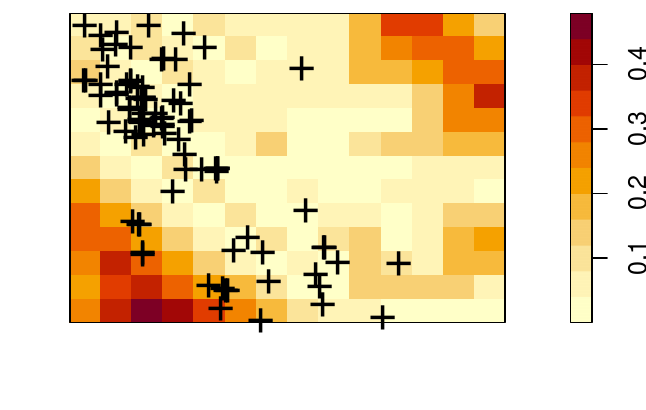}
    \caption{Left panel: Earthquakes occurred in Abruzzo between May 2012 and May 2016, consisting of 85 events
with 2.5 as the threshold magnitude; Right panel: Spatial covariate representing the Distance to the nearest seismic station}
    \label{fig:italy}
\end{figure}

This section addresses the further challenge of estimating the first-order intensity function depending on external covariates. In spatial point process theory, these are known as spatial covariates, and they pose additional challenges with respect to the most common homogeneous or inhomogeneous Poisson processes since, for computational feasibility, their value must be known theoretically at each location.\\
In real data analysis, external covariates, often representing environmental phenomena, are not collected in the same detail as the observed pattern.\\
This implies that the standard quadrature scheme must either be customized to align with the locations of covariates or, conversely, external covariate values must be interpolated at the positions of both data and dummy points. Naturally, this requirement can complicate the implementation of the quadrature scheme method, particularly as the number of covariates increases, considering that each covariate may potentially be gathered at distinct sites.\\
Here we illustrate an example of a spatial point pattern whose realization is assumed to depend on an external covariate, to prove the applicability of our method.
It's important to highlight that a notable advantage of our proposed method is that precise knowledge of covariate values is required only at the data point locations.
The advantage of this approach is its capability to treat both marks and spatial covariates in a consistent manner within the linear predictor of the fitted first-order intensity function.\\
The Italian catalogue considered in this section is downloaded from the Istituto Nazionale di Geofisica e Vulcanologia (INGV) archive. 
As done in \cite{d2022local}, we focus on the seismic sequence of the Abruzzo region. 
Therefore, the analysed earthquakes that occurred between May 2012 and May 2016 in Abruzzo are displayed in the left panel of Figure \ref{fig:italy}, consisting of 85 events with 2.5 as the threshold magnitude. 
On the right panel of Figure \ref{fig:italy}, we display the spatial covariate, whose effect on the intensity of earthquakes we are interested in studying: the distance from the nearest seismic station, henceforth denoted $D_{ns}({u})$.

We therefore proceed by fitting the following model
\begin{equation}
	\label{eq:add0}
	\lambda({u})=\exp(\theta_{0}+\theta_{1}D_{ns}({u})).
\end{equation}
Table \ref{tab:ests_italy} contains the estimates of the model for the Italian earthquake data obtained by both MLE and the MC procedures.

\begin{table}[!ht]
\centering
\caption{Estimates of model for the Italian earthquake data obtained by MLE and MC procedures.}
\begin{tabular}{r|rr|rr}
  \toprule
 & MLE & &MC & \\ 
  & Estimate & S.E.&Estimate& S.E. \\ 
  \midrule
  Intercept & 4.83 &  0.20& 5.36&0.35\\ 
  Distance & -0.08 &0.02 & -0.06 & 0.05 \\ 
   \bottomrule
\end{tabular}
\label{tab:ests_italy}
\end{table}

\begin{figure}[H]
    \centering
\includegraphics[trim={1.25cm 2.25cm .5cm 3.25cm},clip,width=.5\textwidth]{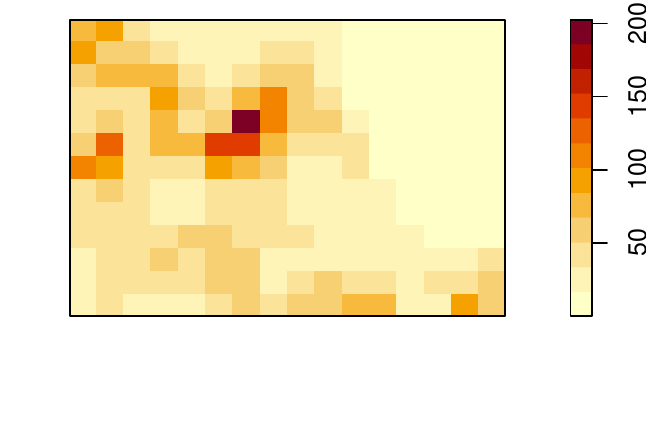}
\includegraphics[trim={0 1.4cm 0 3cm},clip,width=.48\textwidth]{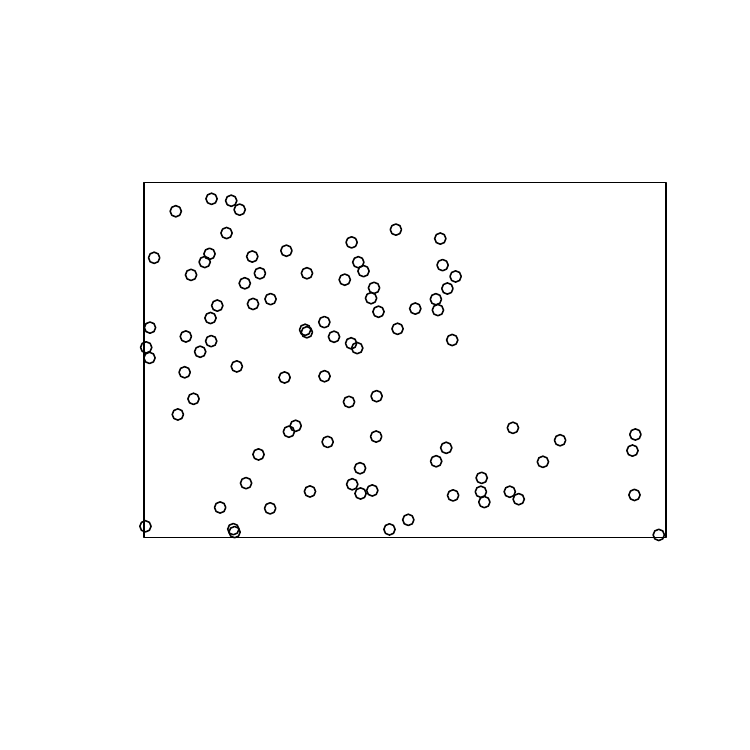}
    \caption{Left panel: intensity function used to simulate the patterns. Right panel: one pattern simulated according to the intensity function.}
    \label{fig:italy2}
\end{figure}

As expected, the MC procedure tends to estimate higher values for the intercept. 
Furthermore, both MLE and MC estimate a negative covariate's coefficient.\\
A negative value of the coefficient of the spatial covariate is reasonable since usually, as the distance from the nearest station recording the event increases, the intensity decreases, as detection of earthquakes is usually more accurate if they occur not far from the network station. This result is in line with those in \cite{d2022local}.\\
To corroborate these results and reinforce the numerical experiments already shown, we run simulations in the presence of an external spatial covariate.\\
In particular, we simulate 100 realizations from the model in equation \eqref{eq:add0}, with the MLE estimated parameters as the true ones, that is, with $\theta_0 =  4.83$ and  $\theta_1 = -0.08$. Figure \ref{fig:italy2} shows the obtained intensity function used to simulate the patterns (left panel) and an example of a simulated pattern (right panel).

Table \ref{tab:simscovs} shows the result for this scenario and also those where the number of points has increased by the double and the quadruple. In these scenarios, the true coefficient of the spatial covariate is kept fixed, and the number of points is increased by adding the logarithm of 2 and 4, respectively, to the true intercept.

\begin{table}[!ht]
\centering
\caption{Mean values over 100 simulations from the model in equation \eqref{eq:add0}, both for MC and MLE methods, for three different numbers of points.}
\begin{tabular}{rr|rrr|rrr|r}
  \toprule
& &  &MC& (R = 0.5)&& MLE&  & $\mathbb{E}[n]$\\ 
& True values &  Estimate &MSE &  S.E. &Estimate & MSE& S.E.&\\ 
  \midrule
Intercept & 4.83&4.81 & 0.03& 0.19   & 4.63&0.08 & 0.34&87.33\\ 
Distance & -0.08&-0.08 &0.0002& 0.02&  -0.06& 0.0006& 0.03&\\ 
  \midrule
Intercept &5.52  & 5.5 &0.02 & 0.14  &  5.45&0.35 &0.35&175.81\\ 
Distance & -0.08  &-0.08&0.0001&0.01 &   -0.08& 0.01& 0.03&\\
  \midrule
Intercept &6.21 &6.19&0.01& 0.10   & 6.29  &0.55 & 0.34&349.11\\ 
Distance & -0.08  & -0.08&0.00008& 0.008   &   -0.09&0.01 &0.03& \\ 
   \bottomrule
\end{tabular}
\label{tab:simscovs}
\end{table}

The MC procedure accurately estimates values close to the true ones. Hence, again, its performance appears to be robust and relatively independent of the sample size.

We note that both methods in equations \eqref{eq:bestR} and \eqref{eq:chooseR}, independently select the same value of $R=0.5$,  providing strong evidence that our selection process is robust and reliable.
\\
A further notice regards the comparison between MC and MLE estimates. The MC estimates exhibit lower bias, with smaller MSE and standard errors when compared to the MLE method. This difference in performance may be due to the application of the radial penalty, with the penalty radius R optimized for this specific case, enhancing the accuracy of the MC estimates compared to the MLE method in this particular scenario.

\subsection{Log-Gaussian Cox processes}

This section is devoted to the assessment of the two-step minimum contrast procedure introduced in section \ref{sec:cox}, specifically tailored for LGCPs.\\
We consider a scenario used in \cite{siino2018joint}, used there to compare the performance with that of the then-proposed joint minimum contrast (JMC) estimation method. \\
The objective is to investigate the possible differences in the estimates of interaction parameters when the summary statistic used in the second stage is weighted by a first-order intensity function estimated through our proposed method.\\
We consider a separable structure for the covariance function of the GRF \citep{brix2001spatiotemporal} that has exponential form for both the spatial and the temporal components,
	$
		\mathbb{C}(r,h)=\sigma^2\exp (-r/\alpha)\exp(-h/\beta),
	$
where $\sigma^2$ is the variance, $\alpha$ is the scale parameter for the spatial distance and $\beta$ is the scale parameter for the temporal one.\\
$200$ point patterns are generated with $n = 1000$ expected number of points in the spatio-temporal window $W \times T = [0,1]^2 \times [0,50]$, with constant first-order intensity equal to $b = \log (n / |W \times T|)$ = 20.\\
We consider a moderate degree of clustering in the processes with variance $\sigma^2 = 5$ and scale parameters in space and time, $\alpha =  0.10$  and $\beta = 2$. The mean of the GRF is fixed $\mu = -0.5 \sigma^2$.\\
Table \ref{tab:lgcp0} reports the estimates of the 200 simulated log-Gaussian Cox processes obtained with the MLE and MC procedure at the first step and the  JMC procedure at the second one.

\begin{table}[!ht]
\caption{Estimates' means and MSE values of 200 simulated log-Gaussian Cox processes with 1000 expected number of points, obtained by both the MLE and MC procedure at the first step, and minimum contrast based on JMC at the second.}
\centering
\begin{tabular}{lr|rr|rr}
  \toprule
& & MLE & &MC & \\ 
 &       True values & Estimate &  MSE & Estimate & MSE   \\ 
  \midrule
$\sigma^2$ & 5  &6.54&5.03  & 6.57 & 3.88  \\ 
$\alpha$&0.05    &0.08&0.001& 0.05 & 0.0002  \\ 
$\beta$&2   &2.43&0.92 &1.79 & 0.40  \\ 
   \bottomrule  
\end{tabular}
\label{tab:lgcp0}
\end{table}

These results are indeed quite promising, especially when considering the second-order parameter estimates obtained using the MC method in the first step. While these estimates differ from the MLE results, they still provide comparable outcomes. Indeed, the variance parameter tends to be overestimated; however, the spatial and temporal range parameter estimates are way more precise compared to those obtained using MLE in the first stage. This improvement can be attributed to the estimated value of the first-order constant intensity function, which has an average of 29.33 when using the MC method, higher than the value of 20 estimated by MLE.

\section{Applications to real data}\label{sec:complex}

This section is dedicated to the analysis of real datasets using the newly proposed inferential framework, with a specific emphasis on the local characteristics of the point patterns.

\subsection{Analysis of copper data}

\begin{figure}[H]
    \centering
    \includegraphics[trim={0 1cm 0 2.5cm},clip,width=.25\textwidth]{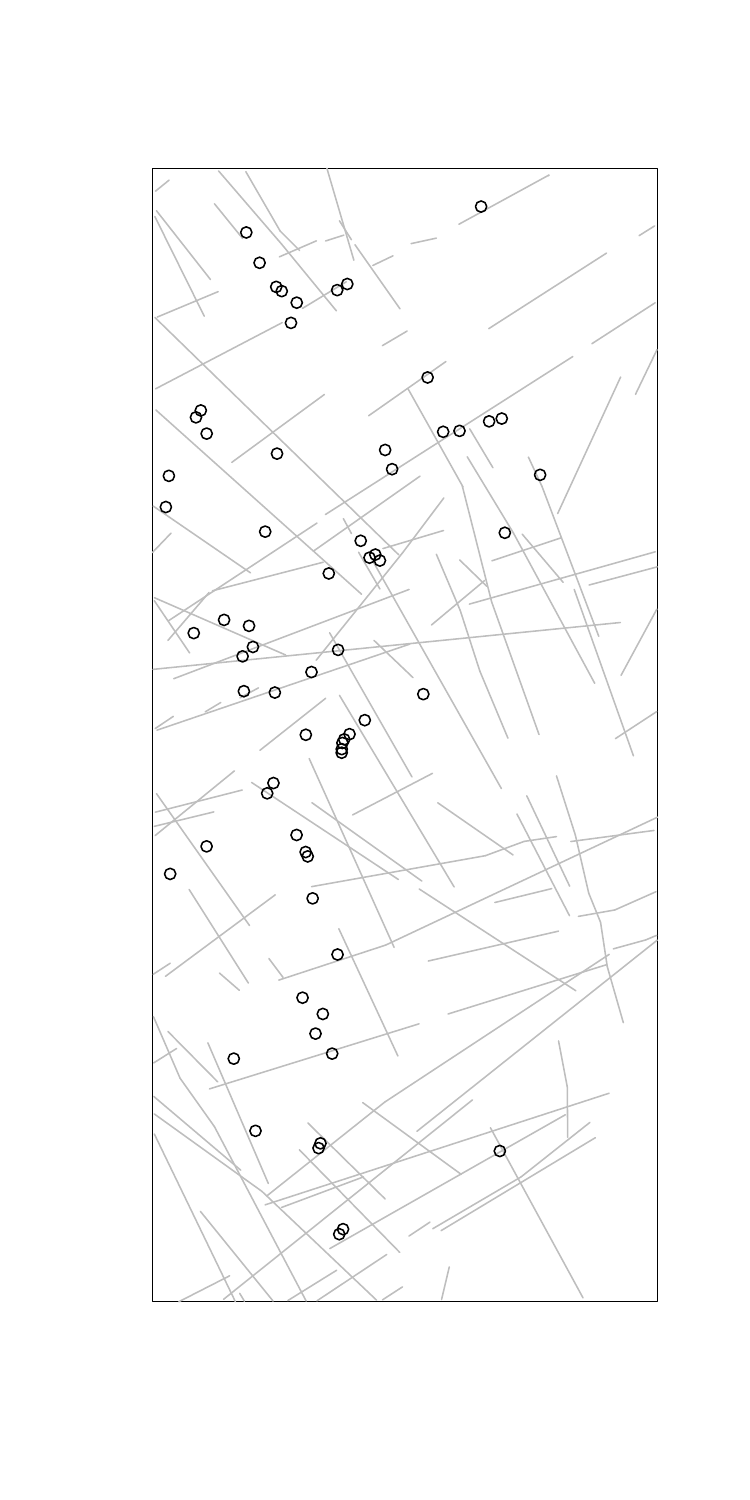}
\includegraphics[trim={0 4.5cm 0 5.5cm},clip,width=.35
\textwidth]{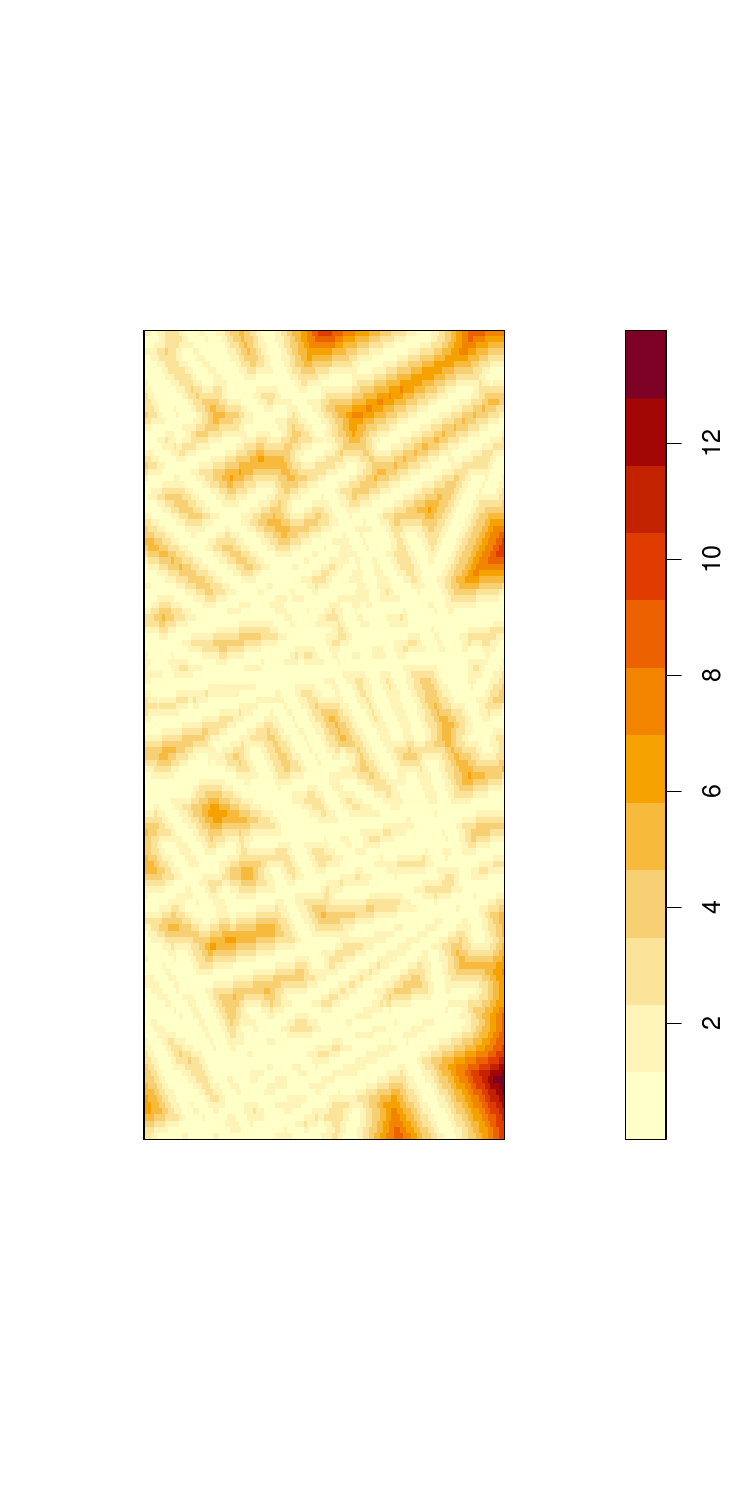}
    \caption{\textit{Left panel}: Berman-Huntington points and lines dataset. Black points are the locations of copper ore deposits, and grey lines are the geological linements. \textit{Right panel}: The available covariate for the copper data: \textit{Distance from the faults}  ($D_f({u})$), computed as the Euclidean distances from the spatial location ${u}$ of events and the   map of geological information.}
    \label{fig:1}
\end{figure}

We examine the Berman-Huntington points and lines dataset, also analyzed by \cite{baddeley:2017local}. The origins and analysis of these data were first presented by \cite{berman:86} and have since been explored by \cite{berman1989estimating, berman1992approximating, baddeley2000practical, foxall2002nonparametric, baddeley2005residual}, to cite a few. \\
These data were collected during an extensive geological survey of a region measuring $70 \times 158$ km in central Queensland, Australia. The dataset comprises 67 points representing copper ore deposits and 146 line segments depicting geological lineaments (left panel of Figure \ref{fig:1}).

Lineaments, visible on satellite imagery, are linear features primarily believed to be geological faults \citep{berman:86}. Typically, the focus lies on predicting copper deposits based on the pattern of these lineaments, which can be readily observed in satellite images.\\
For this reason, we construct the spatial covariate \textit{Distance from the faults} ($D_f$), computed as the Euclidean distances from the spatial location ${u}$ of events and the map of geological information \citep{baddeley:rubak:tuner:15}. The covariate surface is displayed in the right panel of Figure \ref{fig:1}.

We proceed by fitting the following model for the copper ore deposit intensity
\begin{equation}
	\label{eq:add}
	\lambda({u})=\exp(\theta_{0}+\theta_{1}D_f({u})).
\end{equation}

The estimates and their uncertainty are reported in Table \ref{tab:ests}.

\begin{table}[!ht]
\centering
\caption{Estimates of model for the copper data in equation \eqref{eq:add} obtained by MLE and MC procedures.}
\begin{tabular}{r|rr|rr}
  \toprule
 & MLE & &MC & \\ 
  & Estimate & S.E.&Estimate& S.E. \\ 
  \midrule
Intercept & -4.93 &  0.18& -4.52&0.0003 \\ 
  Distance from the faults & -0.10 &0.08 & -0.11 & 0.0001 \\ 
   \bottomrule
\end{tabular}
\label{tab:ests}
\end{table}

The difference between the unpenalized and the penalized procedure is negligible, both in terms of estimates and standard errors, suggesting that the unpenalized procedure suffices for this particular scenario. We want to reiterate the noteworthy similarity between the MC estimates and the MLE estimates.

Finally, we also compare the MC local estimation approach with that of \cite{baddeley:2017local}.
This means to fit the following model, with space-varying parameters
\begin{equation}
	\label{eq:add_local}
	\lambda({u})=\exp(\theta_{0}({u})+\theta_{1}({u})D_f({u})).
\end{equation}
Note that this model differs from the one in equation \eqref{eq:add} since both the parameters are indexed by the spatial location $u$.

\begin{figure}[H]
    \centering
\includegraphics[trim={0 2cm 0 .5cm},clip,width=\textwidth]{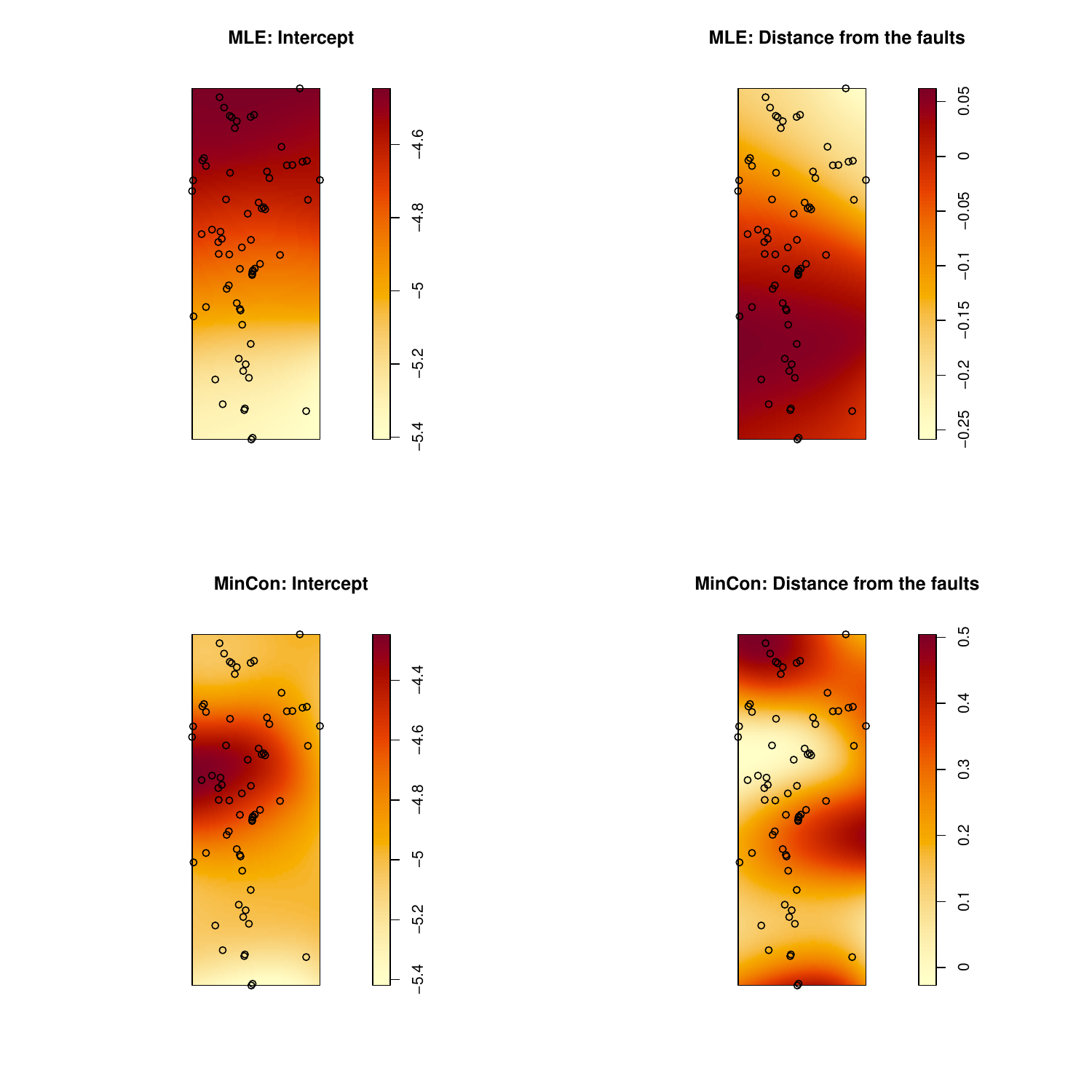}
    \caption{Smoothed local parameters. \textit{Top panels:} MLE estimates. \textit{Bottom panels:} Minimum contrast estimates.}
    \label{fig:3}
\end{figure}

Figure \ref{fig:3} reports the smoothed local estimates for both parameters and both estimation procedures.

These findings highlight the higher variability of the MC estimates in comparison to the MLE estimates. Furthermore, the relatively smoother MLE estimates can be attributed to the underlying kernel technique employed in the log-likelihood function, which is optimized during the MLE process. 
On the other hand, the MC procedure tends to identify more distinct and separated regions in individual point spatial domains. However, it's important to note that a substantial portion of the analyzed region, such as the right-center part, exhibits different values of intensity. This variation is likely attributed to the presence of kernel smoothing artefacts, which can lead to localized inconsistencies in the estimated intensity values.

\subsection{Analysis of Greek seismicity}

For this last scenario, we consider the same data analysed in \cite{d2022local}, related to 1111 earthquakes that occurred in Greece between 2005 and 2014, coming from the Hellenic Unified Seismic Network (H.U.S.N.), with the specific aim of fitting both a global and a local version of Log-Gaussian Cox process (LGCP) model. Indeed, \cite{d2023locally} proposed a local version of spatio-temporal log-Gaussian Cox processes by using Local Indicators of Spatio-Temporal Association (LISTA) functions plugged into the minimum contrast procedure to obtain space as well as time-varying parameters of the covariance structure. The dataset and the functions to fit both global and local LGCPs are available from the \cite{R} package \cite{stopp}.

Our specification involves a constant intensity for the first-order intensity function. Therefore, our primary focus is to investigate how variations in the estimate of the first-order intensity function affects the estimation of the second-order structure, following the approach taken in our previous numerical experiments. Specifically, we utilize the same separable doubly exponential covariance function to ensure a comparable evaluation for our purpose.\\
Table \ref{tab:lgcp} reports the estimates of the log-Gaussian Cox processes applied to the Greek earthquake data with the MLE and MC procedure at the first step.

\begin{table}[!ht]
\caption{Estimates' values (MSE in brackets) of the log-Gaussian Cox processes applied to the Greek earthquake data with the MLE and MC procedure at the first step.}
\centering
\begin{tabular}{r|rr}
  \toprule
& MLE & Min Con\\
& Estimate (S.E.)& Estimate (S.E.)\\
  \midrule
$\lambda$ &    0.006(0.030)    & 0.007(0.000) \\ 
 $\sigma^2$   & 8.15  & 10.95 \\
$\alpha$ & 0.23  &  0.18  \\
$\beta$ & 83.97& 51.69\\
   \bottomrule
\end{tabular}
\label{tab:lgcp}
\end{table}

The covariance parameters are obtained by using the pair-correlation function in the second step, weighted by the first-order intensity function previously estimated.\\
As expected, in the first step, the MC procedure tends to estimate a higher first-order constant intensity function, and this influence is evident in the estimates of the covariance function. The MLE procedure, on the other hand, appears to attribute a greater portion of the process's variability to the clustered structure of the pattern, resulting in smaller variance and higher range parameters. Otherwise, the MC procedure yields an overall higher estimate of the first-order constant intensity function, but less dense clusters with higher variance and smaller range parameters. In other words, the two methods seem to interpret and model the underlying spatial structure differently, with implications for the estimated covariance function and the overall pattern characterization.

\begin{figure}[H]
    \centering
\includegraphics[width=\textwidth]{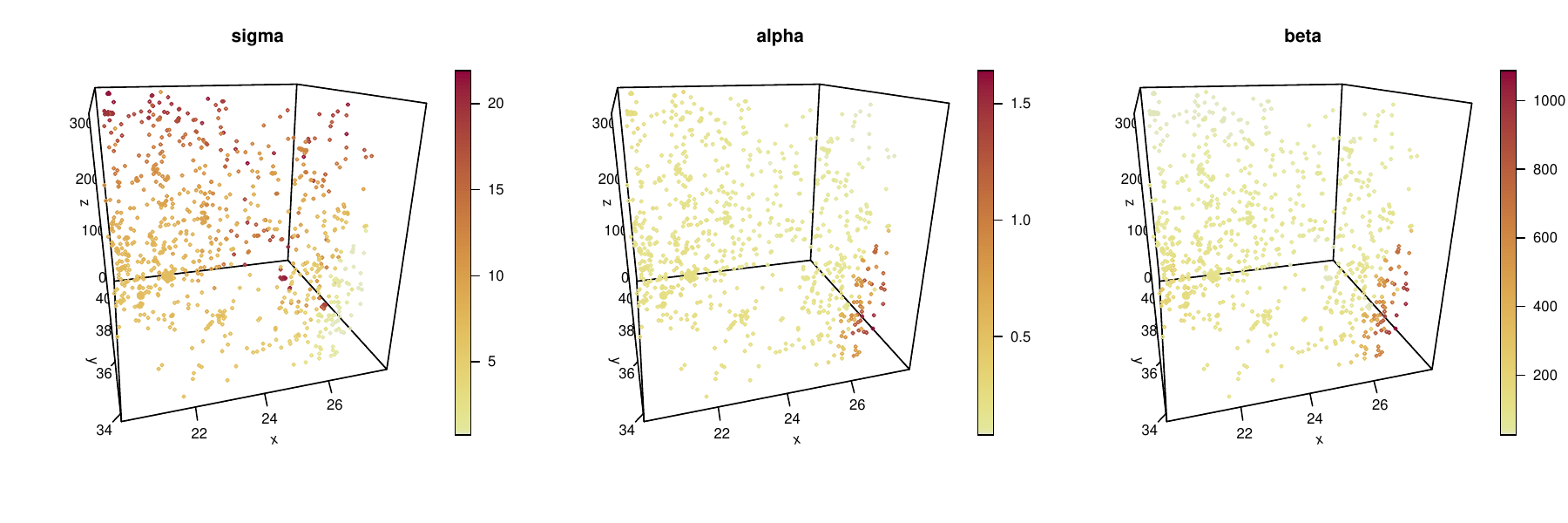}
    \caption{Local estimates of the local LGCP fitted to the Greek seismic data, with the MC method for the estimation of the first-order intensity function. 
    }
    \label{fig:5}
\end{figure}

Note that the magnitude of the parameters, particularly the covariance parameters, depends on spatio-temporal units. Consequently, both MLE and MC procedures may provide overall higher values. In other words, both procedures may lead to the overall conclusion that the observed pattern exhibits clustering characteristics.
Indeed, previous studies showed that this seismic catalog clearly exhibits a clustered structure which can be well described by an LGCP.\\
Furthermore, \cite{d2023locally} showed that a local version of the LGCP further improves the fitting to the data. For this reason, we proceed by fitting a local LGCP to the data, again with both the estimates (MLE and MC) at the first step, and compare the results.\\
With local LGCP, we mean that the first-order intensity function is the same (global one) previously fitted and reported in Table \ref{tab:lgcp}, but we fit individual interaction parameters for each of the points in the observed pattern.
Of course, we also expect these local estimates to change if compared to the ones obtained using MLE at the first step.\\
Figure  \ref{fig:5} shows the result of the two-step minimum contrast fitting.

As expected, the local estimates slightly change between the two methods.
Indeed the conclusions one could draw from this application could be the same as in \cite{d2023locally}, meaning that both estimations permit to observe regions where points tend to cluster in smaller clusters (bottom right area: small variance and high range parameters) and other where clustering is more mild.\\
However, we still identify some differences, particularly regarding the MC selecting a smaller area with the most dense clustering.

\section{Conclusions and future work}\label{sec:conclusions}
This paper presents a novel fitting procedure for the first-order intensity function of point processes, based on the minimum contrast procedure.
Knowing the expectation of the $K$-function when weighted by the true intensity function, we utilize this result to construct a model estimation procedure that is adaptable to any model specification. The only prerequisite is the knowledge of the expression of the first-order intensity function, completely circumventing the need to face the likelihood, which can often be complex to maximize, in point process models. 

The motivation behind our research stems from the desire to simplify the estimation process and broaden its applicability. By employing the expectation of the weighted $K$-function, our method offers an intuitive solution to the complexities associated with point process models. 
In detail, compared to MLE and standard point process estimation methods, our method completely avoids dealing with the likelihood and, therefore, avoids the complexities of its two- or three-dimensional integrals.
Turning to the local estimation context, our method is able to provide local estimates without introducing any additional complexity, than the global estimation.

This paper particularly deals with Poisson process models, whose likelihood represents the foundation of many more complex models.

We have presented simulated results for both purely spatial and spatio-temporal contexts, along with the analysis of two real datasets related to environmental issues. These real datasets serve to illustrate more complex scenarios than the Poisson one.
An important finding from our simulation studies is that our approach appears to be robust to variations in sample size, whether in spatial or spatio-temporal patterns. This finding is particularly relevant as it suggests that our method may not require exceptionally large datasets to yield reliable results.

In conclusion, our approach opens the path for future research utilizing the minimum contrast procedure for first-order intensity estimation in considerably more complex models than the Poisson model, as already sketched in this study.

It is important to stress that the results presented in this paper concerning the Poisson case are highly noteworthy. Indeed, in our approach, we have entirely overcome the need for likelihood maximization and the subsequent approximation of the integral over the intensity function. Furthermore, we have not required any quadrature scheme or the selection of kernel weights for local estimation. Despite these simplifications, our method has yielded results that are comparable to those obtained through Maximum Likelihood Estimation. This highlights the effectiveness and validity of the proposed approach, which can significantly reduce the computational complexity, maintaining accuracy and reliability in parameter estimation, and producing results that do not deviate much from the more traditional MLE approach.

Many research paths could be deepened in future.
First, we believe that the optimization procedure could be improved by weighting the objective function in equation   \eqref{eq:min_con}  by a $\phi(r,h)$ function. This would basically correspond to giving more importance to some specific spatial and temporal lags.
For instance, \cite{diggle2013statistical} suggested using $\phi(r,h)$ to weight the discrepancy measure by the inverse (approximate) variance of the $K$-function (in which \cite{guan2007least}, used their sub-sampling method to achieve). 
The variance of the $K$-function, however,  is typically unknown.  For a spatial Poisson process,  it is suggested to use $\phi(r) =r^{-2}$, but no other specific recommendations for other types of process are given.
Then, we want to run extended simulation studies to assess the performance of the proposed procedure in more complex settings, for instance, with Self-Exciting models such as the ETAS ones.

Furthermore, in parallel to our idea, \cite{kresinestimation} showed that parameters in spatio-temporal point process models, alternatively to MLE, can be estimated consistently,
under general conditions by minimizing the Stoyan-Grabarnik statistic.
Therefore, we wish to compare our proposal based on the $K$-function to 
\cite{kresinestimation}'s proposal.

\section*{Funding}
This work was supported by the Targeted Research Funds 2024 (FFR 2024) of the University of Palermo (Italy), the Mobilità e Formazione Internazionali - Miur INT project ``Sviluppo di metodologie per processi di punto spazio-temporali marcati funzionali per la previsione probabilistica dei terremoti", and by the European Union -  NextGenerationEU, in the framework of the GRINS -Growing Resilient, INclusive and Sustainable project (GRINS PE00000018 – CUP  C93C22005270001). The views and opinions expressed are solely those of the authors and do not necessarily reflect those of the European Union, nor can the European Union be held responsible for them.

\bibliography{AAA}

\end{document}